\documentclass{kluwer}    % Specifies the document style.
\input psfig.tex 
\newdisplay{guess}{Conjecture}

\begin{document}
\begin{article}
\begin{opening}
\title{Gas in Groups and Clusters of Galaxies}
\author{Sabine \surname{Schindler}}
\runningauthor{Sabine Schindler} \runningtitle{Gas in Groups and Clusters}
\institute{Institut f\"ur Astrophysik, Universit\"at Innsbruck}
\date{}

\begin{abstract}
Groups and clusters contain a large fraction of hot gas which emits
X-ray radiation. This gas yields information on the dynamical state
and on the total mass of these systems. X-ray spectra show that heavy
elements are present in the gas. As these metals must have been
produced in the cluster/group galaxies and later transported into the
gas, the metallicity is a good tracer for the transport processes. Several
possible processes, that transport gas from the small potential wells
of the galaxies into the clusters and groups, are discussed.
\end{abstract}

\end{opening}

\section{Gas properties}

Clusters and groups of galaxies do not only contain galaxies but all
the space between the galaxies is filled with hot gas. This gas is
also often referred to as intra-cluster medium (ICM) or
intra-group medium (IGM). The high temperature of the gas of 1 -10 keV
is in
good correspondence with the depth of the potential wells of
groups and clusters (see Table 1). According to the shallower
potential of groups also the temperature is lower in groups ($\approx$
1 keV) compared to clusters (3-10 keV). The
gas is almost fully ionised and has very low densities of $10^{-2}
- 10^{-4}$ cm$^{-3}$ decreasing outwards. It is optically
thin. In the spectra of both, groups and clusters, lines of heavy
elements have been detected corresponding to metallicities of
about 0.2 to 0.4 in solar units. These metals indicate that the
gas cannot be of purely primordial origin, but part of it must
have been previously the inter-stellar medium (ISM) in galaxies and
then it must have been transported from the galaxies into the ICM and IGM,
respectively, by certain processes. The processes will be
discussed in the next section.

\begin{table} %
\begin{tabular}{lll}
\hline
              & Clusters  & Groups\\
\hline
Temperature   & 3-10 keV       & $\approx 1$ keV\\
Extent        & few Mpc        & 0.1-1 Mpc\\
Metallicity   & 0.2-0.4 solar  & 0.2 solar\\
X-ray luminosity & $10^{43-45}$ erg/s & $10^{42-43}$ erg/s\\
\hline
\end{tabular}
\caption[]{Properties of the gas in cluster and groups of galaxies.}
\label{parset}
\end{table}

Gas with such properties emits thermal bremsstrahlung in the X-ray
range (see Table 1). Therefore groups and clusters are currently
studied extensively by the X-ray observatories XMM and CHANDRA.
While in almost all of the clusters X-ray emission has been found,
the X-ray emission from groups is somewhat harder to detect due to
their lower luminosity. So far in about 50\% of the nearby groups
X-ray emission has been found.

\begin{figure} %
\psfig{figure=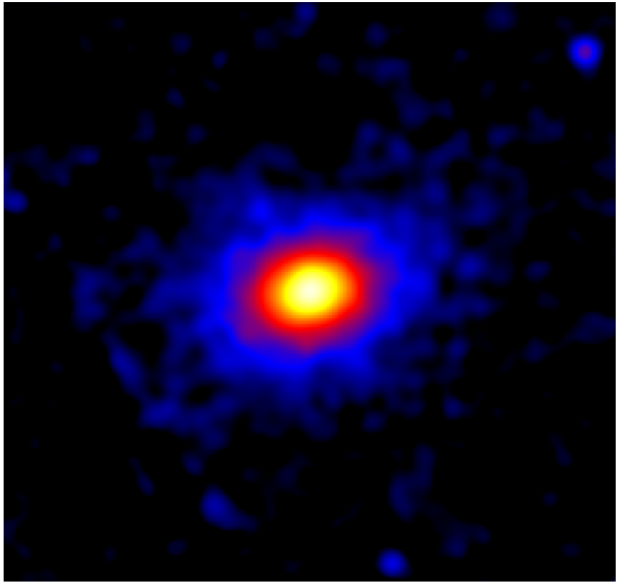,width=8.2cm,clip=}
\caption[]{CHANDRA image of the cluster RBS797 (from Schindler et
al. 2001). The cluster looks well virialised with a very regular surface
brightness distribution.}
\label{RBS797}
\end{figure}

\section{Gas distribution, dynamical state and mass determination}

Two examples of X-ray emission from
clusters of galaxies are shown in Figs.~1 and 2. 
The cluster RBS797 (Fig.~1) has a very regular
shape on this scale with a somewhat elongated structure. This is a
typical relaxed cluster.
The cluster CL0939+4713 (Fig.~2) has a very
different, irregular morphology. In the XMM image one can
distinguish clearly two main subclusters. These subclusters show
even some internal structure. Obviously the cluster is not in an
equilibrium state, but there are still some structures visible
that are in the process of falling into the cluster. So already
the X-ray morphology contains a lot of information about the dynamical
state of the cluster. But X-ray spectroscopy tells even more. In
Fig. 3 the temperature distribution of the gas in CL0939+4713 is shown. There is
a hot region between the two subclusters which is an indication
that the subclusters are approaching each other -- just as it is
expected from hydrodynamic simulations (Schindler \& M\"uller 1993;
Ricker \& Sarazin 2001; Ritchie \& Thomas 2002): when two subclusters 
start to merge the gas between them is compressed and heated,
so that a hot region is produced. Therefore we conclude that this
cluster in the beginning of a major merger process (De Filippis et
al. 2002).

\begin{figure} %
\psfig{figure=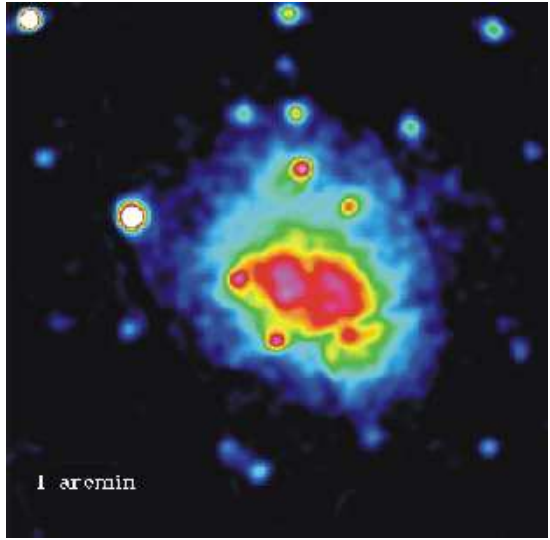,width=8.2cm,clip=}
\caption[]{XMM image of the cluster CL0939+4713 (from De Filippis et
al. 2002). Two subclusters are visible with even some internal
structure indicating that the cluster is in a merger process.}
\label{}
\end{figure}

\begin{figure} %
\psfig{figure=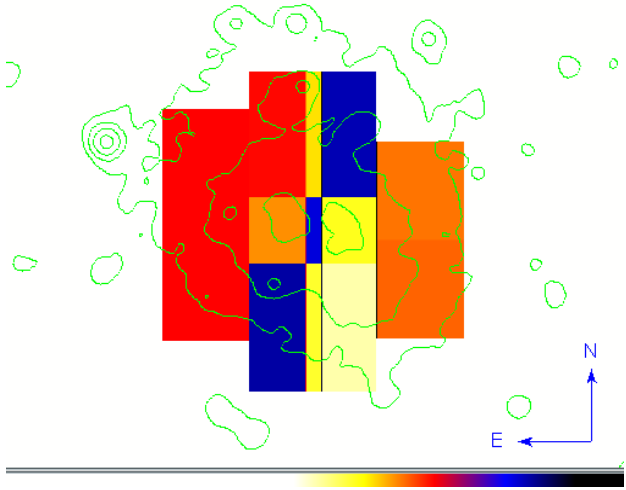,width=8.2cm,clip=}
\caption[]{Temperature map of the cluster CL0939+4713 (from De Filippis et
al. 2002). The temperature is increasing 
%with increasing darkness of the grey scale.
from yellow over orange and red to blue. 
The contours show the same surface brightness
distribution as in Fig.~2. Between the two subclusters a
region of hot gas is visible indicating that the subclusters are approaching
each other.}
\label{}
\end{figure}

Also groups show extended emission coming from the IGM as it can be
seen e.g. in HCG16 (Belsole et al. 2002)  or in HCG90 (Longo et al. 2002).
In groups the extended emission from the IGM is not so dominant as in
clusters. One can usually see different components in the X-ray
emission. Apart from the extended IGM also more
compact components caused by galaxies and other sources can be seen.

The mass of the ICM and IGM is not negligible compared to the
other components. Table 2 gives an overview of the mass fraction
of gas, galaxies and dark matter in groups and clusters,
respectively. In clusters there is always considerable more mass
in the gas than in the galaxies, while in groups the mass in the
galaxies usually exceeds the mass in the gas. Both systems have in
common that most of the mass is in form of dark matter. This shows
how important mass determination is. When the total mass of the
system can be measured and the visible components are subtracted,
the amount and distribution of the dark matter can be inferred.

\begin{table} %
\begin{tabular}{lll}
\hline
              & Clusters  & Groups\\
\hline
Galaxies    & 3-5\%       & 3-20\%\\
Gas        & 15-20\%        & 2-10\%\\
Dark matter   & rest  & rest\\
\hline
\end{tabular}
\caption[]{Mass fraction of the different components in clusters
and groups of galaxies
}\label{parset}
\end{table}

There are different ways to determine masses, one of them is the
X-ray method. With two assumptions -- spherical symmetry and
hydrostatic equilibrium -- the X-ray emitting gas can be used as a
tracer for the total potential. Only the gas density and the gas
temperature are required for this method, which can both be measured from X-ray
observations.

With the gas density also the gas mass is known. The ratio between
gas mass and total mass is the gas mass fraction. In Fig.~4 the
gas mass fraction is plotted versus the radius. We find that the
gas mass fraction is not constant but is increasing with radius
(Schindler 1999; Castillo-Morales \& Schindler 2002) .
That means that the gas is more extended than the dark matter.
Obviously, there is not only gravitational energy from the
collapse present but there must be additional heating processes at
work like
e.g. supernova-driven winds. Also interesting is the question 
whether the gas mass
fraction is changing with time. In Fig.~5 the
gas mass fraction is shown for nearby and distant cluster samples. So far
no significant evolution is detectable.

\begin{figure} %
\psfig{figure=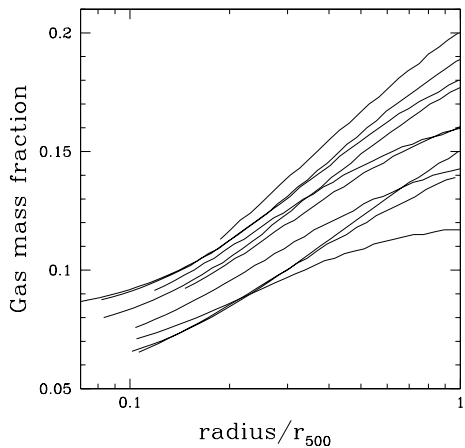,width=7.2cm,clip=}
\caption[]{Gas mass fraction versus radius. For all the clusters the
gas mass fraction increases with radius, i.e. the gas distribution is
more extended than the dark matter distribution (from Castillo-Morales
\& Schindler 2002).}
\label{RBS797}
\end{figure}

\begin{figure} %
\psfig{figure=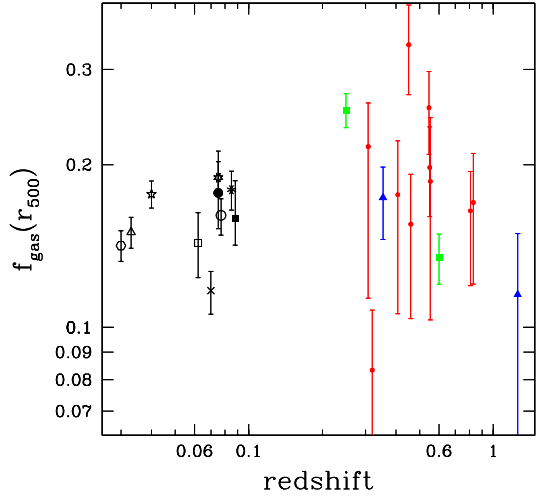,width=8.2cm,clip=}
\caption[]{Gas mass fraction versus redshift. With the current data
not significant evolution of the gas mass fraction in clusters is visible 
(from Castillo-Morales
\& Schindler 2002).}
\label{RBS797}
\end{figure}

The gas mass fraction can also be used to derive an upper limit on
the matter density of the universe $\Omega_m$. Assuming that the
matter accumulated in a cluster is representative for the universe
as a whole, the ratio between the mass in baryons and the total
mass $M_{gas}/M_{tot}\approx f_{gas}$ should be the same as
$\Omega_{baryon}/\Omega_m$. As an upper limit for 
$\Omega_{baryon}$ can be determined
from primordial nucleosynthesis to be about 0.06 (for $H_0 = 50$
km/s/Mpc) one finds $\Omega_m \approx 0.3 - 0.4$ in units of the critical
density.

\section{Interaction between ISM and ICM/IGM}

The ICM and the IGM show metal lines in the X-ray spectra.
These metals cannot have been produced in the gas, but they must have
been produced in the galaxies and subsequently transported from
the galaxies into the ICM/IGM by certain processes like
e.g. ram-pressure stripping, galactic winds, galaxy-galaxy interaction
or jets from active galaxies.
The metallicity is the best indicator for finding out which of these 
processes are most important. Of special
interest is the distribution of metals. So far there are only
few examples of measured metallicity variations in real 2D
maps and not only profiles. In CL0939+4713 we find different
metallicity in the different subclusters (De Filippis et al.
2002). In the Perseus cluster also clear metallicity variations were
found (Schmidt et al. 2002). 1D profiles are not very useful in this
context because photons from regions in the cluster which are very
far apart are accumulated in the same spectrum.

Apart from the metallicity distribution also the evolution of the
metallicity is
interesting. As soon as enough XMM and CHANDRA observations of
distant clusters are available we can compare the metallicities in
these clusters with those of nearby clusters. This is another way of
distinguishing between the enrichment processes as different processes
have different time dependence.
In addition element
ratios can be derived, e.g of Fe and $\alpha$-elements to get
information on the different  types of 
supernovae that have contributed to the
metal enrichment.

Various processes have been suggested for the transport of gas
from the galaxies to the ICM/IGM. 30 years ago Gunn \& Gott (1972)
suggested ram-pressure stripping: as the galaxy moves through the
cluster and approaches the
cluster centre it feels the increasing pressure of the
intra-cluster gas. At some point the galaxy is not able anymore
to retain its ISM. The ISM is stripped off and lost to the ICM and
with it all its metals. Many numerical simulations have
been performed to investigate this process, first 2D models (Takeda
et al. 1984; Gaetz et al. 1987; Portnoy et al. 1993; Balsara et
al. 1994). With increasing computing power also more detailed 3D
models could be calculated (Abadi et al. 1999; Quilis et al. 2000;
Vollmer et al. 2001; Schulz \& Struck 2001; Toniazzo \& Schindler
2001). In Fig.~6 such a simulated stripping process is shown for an
elliptical galaxy.

\begin{figure} %
\psfig{figure=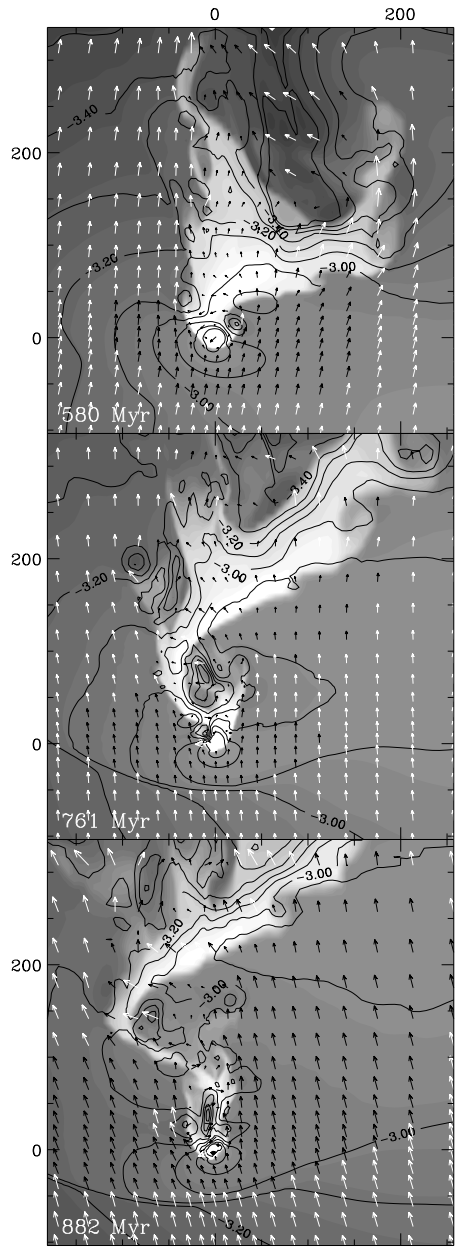,width=6cm,clip=}
\caption[]{Gas density (grey scale) and pressure (contours) of a galaxy
moving downwards towards the cluster centre. The arrows show the Mach
vectors (white when $M>1$, black otherwise). The gas of the galaxy is
stripped due to ram pressure (from Toniazzo \&
Schindler 2001).}
\label{RBS797}
\end{figure}

Another possible process is galactic winds e.g. driven by
supernovae (De Young 1978). Also for this process simulations have
been performed on order to see whether only winds can account for
the observed metallicities. The results were quite discordant as
the following two examples show. Metzler \& Evrard (1994, 1997) found
that winds can account for the metals, while Murakami \& Babul (1999)
concluded that winds are not very efficient for the metal enrichment.
In the simulations of Metzler \& Evrard quite steep metallicity
gradients showed up which are not in agreement with observations.

A third possible process is galaxy-galaxy interactions, like
tidal stripping or galaxy harassment. Also during these
interactions a lot of ISM can be lost to the ICM and IGM. This
process is very likely more efficient in groups of galaxies,
because in these systems the relative velocities are smaller and
therefore the interaction timescales are longer. The ram-pressure
stripping on the other hand is probably less efficient in groups
because not only the pressure of the IGM is lower than that of the
ICM, but also the velocities are smaller. This is also very important
as the stripping is about
proportional to $\rho_{gas} v^2$.

A forth possible mechanism is jets emitted by active galaxies.
These jets can also carry metals. Fig.~7 shows the interaction of
jets with the ICM as it was discovered by X-ray observations. In
the cluster RBS797 minima in the X-ray emission have been detected
in a CHANDRA observation (Schindler et al. 2001). The X-ray
depressions are arranged opposite with respect to the cluster
centre. It is very likely that the pressure of the relativistic
particles in the jets push away the X-ray gas. Preliminary radio
observations with the VLA confirm this hypothesis.

\begin{figure} %
\psfig{figure=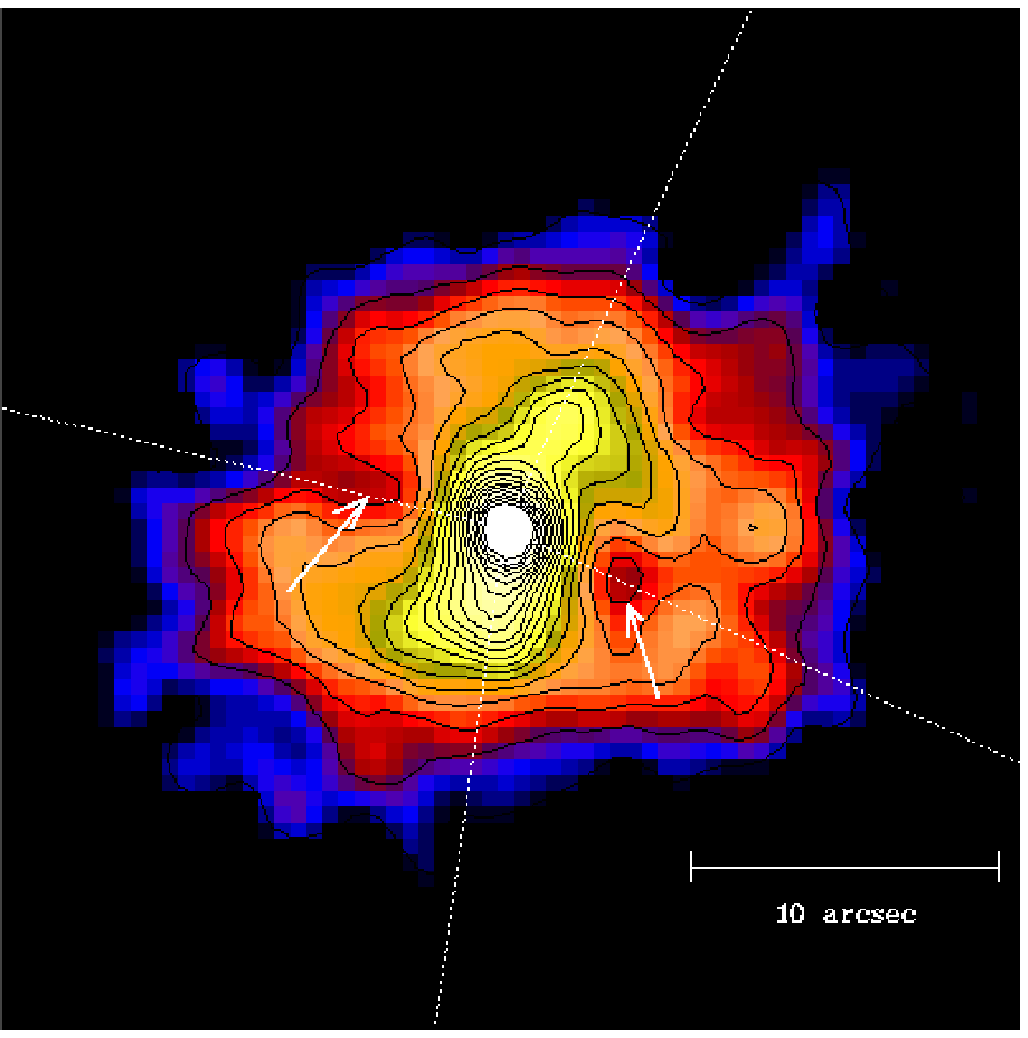,width=7.6cm,clip=}
\caption[]{CHANDRA image of the central part of the cluster RBS797
(from Schindler et al. 2001). There are depressions in the X-ray
emission which are located opposite to each other with respect to the
cluster centre (see arrows). These depressions can be explained by an
active galaxy in the centre of the cluster, which has two jets. The
pressure of the relativistic particles in the jets push away the X-ray
gas resulting in minima in the X-ray emission. }
\label{RBS797}
\end{figure}

Simulations with different enrichment processes were also
performed on cosmological scales. Also here quite discordant
results have been found as the two following examples show. Gnedin
(1998) found that galactic winds play only a minor role, while galaxy
mergers eject most of the gas. In contrary to these results Aguirre
et al. (2001) concluded that winds are most important and ram-pressure
stripping is not very efficient. The reason for these
differences are probably the large ranges in scale that are
covered by these simulations, from cosmological scale down to galaxy
scales. Therefore only a small number of
particles are left for each single galaxy and hence galaxies are not
well resolved. This can be the reason for the discordant results.

In order to clarify this we are currently performing comprehensive
simulations, which include the different enrichment processes.

\section{Summary}

So far the picture of the
interaction between ICM/IGM and the ISM is not very clear. 
There are many different processes possible that can transport gas from
the galaxies to the ICM/IGM:
ram-pressure stripping, galactic winds, galaxy-galaxy interaction,
jets from active galaxies and maybe even more. 
It is not known yet, which of the
processes is most efficient, what the time variations are and how
much it depends on other parameters like e.g. 
the mass of the cluster/group. The best
property to test these different processes is the metallicity. The
distribution and the time evolution of the metals are now
finally measurable with high accuracy with the X-ray
satellites XMM and CHANDRA, so that we will soon get answers on the
importance of
the different enrichment processes.

\acknowledgements
We are very grateful to Bernd Aschenbach for making the XMM data of
CL0939+4713 available to us. They are part of the TS/MPE guaranteed time.

% The endnotes section will be placed here.

\theendnotes

\end{article}
\end{document}